\documentclass[conference]{IEEEtran}

\usepackage{cite}
\usepackage{amsmath,amssymb,amsfonts}
\usepackage{graphicx}
\usepackage{xcolor}
\usepackage{array,multirow}
\usepackage{tablefootnote}
\usepackage{threeparttable}
\usepackage{booktabs, makecell, tabularx}
\usepackage[caption=false, font=footnotesize]{subfig}
\usepackage{dblfloatfix}
\usepackage[hidelinks,bookmarks=false]{hyperref}

\def\BibTeX{{\rm B\kern-.05em{\sc i\kern-.025em b}\kern-.08em
    T\kern-.1667em\lower.7ex\hbox{E}\kern-.125emX}}

\hypersetup{pdfstartview={XYZ null null 1.00}}

\makeatletter
\IEEEtriggercmd{\reset@font\normalfont\fontsize{7.9pt}{8.60pt}\selectfont}
\newcommand*\titleheader[1]{\gdef\@titleheader{#1}}
\AtBeginDocument{%
  \let\st@red@title\@title
  \def\@title{%
    \bgroup\normalfont\normalsize\centering\@titleheader\par\egroup
    \vskip1ex\st@red@title}
}
\makeatother
\IEEEtriggeratref{1}

\titleheader{\vspace{-30pt}To appear at the 29th IEEE International Workshop on Computer Aided Modeling and Design of Communication Links\\and Networks (CAMAD'24), 21-23 Oct, 2024, Athens, Greece.\vspace{-6pt}}
\title{\hspace{-9pt}Towards \hspace{-8pt} Enabling \hspace{-8pt} 5G-NTN \hspace{-8pt} Satellite \hspace{-8pt} Communications\\for Manned and Unmanned Rotary Wing Aircraft}

\makeatletter
\def\ps@IEEEtitlepagestyle{
  \def\@oddfoot{\mycopyrightnotice}
  \def\@evenfoot{}
}
\def\mycopyrightnotice{
  {\footnotesize
  \begin{minipage}{\textwidth}
  \centering%
  ~\copyright~2024 IEEE.  Personal use of this material is permitted.  Permission from IEEE must be obtained for all other uses, in any current or future media, including reprinting/republishing this material for advertising or promotional purposes, creating new collective works, for resale or redistribution\\to servers or lists, or reuse of any copyrighted component of this work in other works.
  \end{minipage}
  }
}

\begin{document}

\author{\IEEEauthorblockN{%
Vasileios Leon\IEEEauthorrefmark{1},
Ilias Christofilos\IEEEauthorrefmark{1},
Athanasios Nesiadis\IEEEauthorrefmark{1},
Iosif Paraskevas\IEEEauthorrefmark{1},\\[2pt]  
Juan Perrela\IEEEauthorrefmark{2},
Georgios Ioannopoulos\IEEEauthorrefmark{3},
Alexandros Tasoulis--Nonikas\IEEEauthorrefmark{3},
Mathieu Bernou\IEEEauthorrefmark{3},
Jacques Reading\IEEEauthorrefmark{4}} \\[-9pt] 
\IEEEauthorblockA{\IEEEauthorrefmark{1}\emph{Intracom Defense S.A., 21st Km Markopoulou Ave., 19441 Koropi, Greece}}
\IEEEauthorblockA{\IEEEauthorrefmark{2}\emph{Alpha Unmanned Systems, Fuente Nueva Ave., 14 Nave 16A, 28703 Madrid, Spain}}
\IEEEauthorblockA{\IEEEauthorrefmark{3}\emph{OHB Hellas, Imvrou 1, 15124 Marousi, Greece}}
\IEEEauthorblockA{\IEEEauthorrefmark{4}\emph{European Space Agency, Keplerlaan 1, 2201 AZ Noordwijk, Netherlands}}\\[-10pt] 
\IEEEauthorblockA{%
\fontsize{9.5}{10}\selectfont
Emails:
\IEEEauthorrefmark{1}\{vleon, iparaskevas\}@intracomdefense.com, \IEEEauthorrefmark{2}jperrela@alphaunmannedsystems.com, 
\\[1pt] 
\IEEEauthorrefmark{3}mathieu.bernou@ohb-hellas.gr,
\IEEEauthorrefmark{4}jacques.reading@ext.esa.int}} 

\maketitle

\begin{abstract}
Satellite Communications (SatCom) are a backbone of worldwide development. In contrast with the past, when the GEO satellites were the only means for such connectivity, nowadays the multi-orbital connectivity is emerging, especially with the use of satellite constellations. Simultaneously, SatCom enabled the so-called In-Flight Connectivity, while with the advent of 5G-NTN, the development of this market is being accelerated. However, there are still various missing points before such a technology becomes mainstream, especially in the case of Rotary Wing Aircraft (RWA). Indeed, due to their particular characteristics, such as the low altitude flights and the blade interference, there are still open challenges. In this work, an End-to-End (E2E) analysis for the performance of SatCom under 5G-NTN for manned and unmanned RWA is performed. Various scenarios are examined, and related requirements are shown. The effects of blades and other characteristics of the RWA are established, and simulations for these cases are developed. Results along with related discussion are presented, while future directions for development are suggested. This work is part of the ESA ACROSS-AIR project.
\end{abstract}


\section{Introduction}
Over the last years, Satellite Communications (SatCom) \cite{satcom_survey} have attracted significant attention by several markets. 
This renewed interest is motivated by the innovative technological advancements and developments that have taken place in the industries of space, computing systems, and telecommunications. 
Regarding the space segment \cite{newspace}, 
the growth of Low Earth Orbit (LEO) constellations, 
the satellite miniaturization, 
the decreased cost of satellite launch, 
and the use of higher frequency bands,
to name a few, 
have brought SatCom to the fore. 
At the same time, the evolution of Internet of Things (IoT), 
facilitated by the emergence of novel edge processors \cite{space_computing},
along with the development of next-gen networks
(5G, B5G, and soon 6G) \cite{5gntn_review},
have marked a new era for SatCom. 

In this evolved landscape, 
SatCom have found a plethora of applications, such as media broadcasting, internet services, data gathering, and mobile communications.   
Apart from the traditional markets,
SatCom have also gained momentum in maritime, aviation, defense and security,
in an effort to improve key aspects of the society. 
These markets are favored by broadband SatCom,
which pave the way 
for services such as:
real-time Earth Observation (EO), remote sensing and monitoring, 
broadband coverage to underserved areas 
(e.g., remote, rural, maritime),
and network availability during an emergency situation or natural disaster. 

Among the affected markets,
the aviation industry supports novel scenarios via SatCom,
especially with modern types of aircraft.
The ever-growing use of  
Unmanned Aerial Vehicles (UAVs) \cite{uav_survey}
in civilian applications,
such as 
infrastructure inspection,
precision agriculture, 
Search \& Rescue (S\&R),
goods delivery,
and 
wildfire detection,
can be enhanced by 
the increased network coverage, 
availability and reliability
of SatCom
compared to terrestrial networks.
The same applies to 
Urban Air Mobility (UAM) \cite{uam_survey}  
vehicles, 
which offer 
intelligent transportation
and immediate response to emergency situations. 
Nevertheless, 
it is still under investigation 
if technological advancements, 
such as LEO mega constellations, 5G networks, and novel antenna designs, 
can provide sufficient data rates, error performance, and quality of service in these scenarios.  
The efficient deployment of SatCom is also 
challenged by aircraft constraints 
(e.g., blades' interference, limited space for antennas and modems, and restricted power budget). 

\begin{table*}[bp]
\vspace{-4pt}
\renewcommand{\arraystretch}{1.07}
\caption{Real-World Scenarios for Rotary Wing Aircraft with Satellite Communications}
\vspace{-3pt}
\label{tb_scenarios}
\setlength{\tabcolsep}{6.4pt}
\centering
\begin{threeparttable}
\begin{tabular}{>{\centering\arraybackslash}m{1cm} >{\centering\arraybackslash}m{1cm}  >{\centering\arraybackslash}m{3cm}   >{\centering\arraybackslash}m{3cm}  >{\centering\arraybackslash}m{3cm}  >{\centering\arraybackslash}m{1.5cm} c}
\hline 
\textbf{Aircraft} & \textbf{Scenario} & & \multicolumn{1}{c}{\textbf{Operational}} &  & \textbf{Data Rate} & \textbf{Service Availability} \\[-1.8pt]
\textbf{Type} & \textbf{ID} & \multicolumn{1}{c}{\textbf{Scenario Mission}} & \multicolumn{1}{c}{\textbf{Environment}} & \multicolumn{1}{c}{\textbf{Data Types}} & \textbf{Requirement{\tnote{1}}} & \textbf{Requirement{\tnote{2}}} \\
\hline
\hline \\[-6.5pt]
\parbox[t]{10mm}{\multirow{18}{*}{\rotatebox[origin=c]{45}{\hspace{5pt}UAV}}} 
& 1 & Industrial Infrastructure Inspection    & industrial zones, tunnels, roads, factories & visual, thermal, multispectral, LiDAR  & high & high\\
\cmidrule(lr){2-7}
& 2 & Physical Network Surveillance           & electricity powerlines, railways            & visual, thermal, multispectral         & high & high\\
\cmidrule(lr){2-7}
& 3 & Mining and Construction Management      & mine \& construction sites, tunnels, roads  & visual, magnetic, multispectral, LiDAR & high & medium\\
\cmidrule(lr){2-7}
& 4 & Livestock and Agricultural Management   & farms, meadows, \hspace{11pt} hills, barns                & visual, thermal                        & medium & low\\
\cmidrule(lr){2-7}
& 5 & Medical Supplies Delivery               & islands, forests, deserts, mountains        & command \& control                     & low & high\\
\cmidrule(lr){2-7}
& 6 & Fire Recognition and Surveillance       & mountains, forests, industrial zones        & visual, thermal, multispectral         & high & high\\
\cmidrule(lr){2-7}
& 7 & Forest and Maritime Observation         & forests, mountains, \hspace{11pt} seas,  lakes             & visual, thermal, multispectral, LiDAR  & high & high\\
\cmidrule(lr){2-7}
& 8 & Precise and Remote \hspace{1pt} Data Gathering       & urban \& rural areas \hspace{11pt} of interest            & visual, thermal, multispectral, LiDAR  & medium & low\\[6pt]
\hline \\[-6.5pt]
\parbox[t]{12mm}{\multirow{8.5}{*}{\rotatebox[origin=c]{45}{\hspace{5pt}UAM}}} 
& 9 & Urban Aerial Transportation             & cities, airports                            & visual, audio                          & high & medium\\
\cmidrule(lr){2-7}
& 10 & Suburban Aerial Transportation         & suburbs, villages, airports                 & visual, audio                          & high & medium\\
\cmidrule(lr){2-7}
& 11 & Aerial Ambulance                       & cities, urban \& rural remote areas         & visual, audio                          & high & high\\
\cmidrule(lr){2-7}
& 12 & Aerial Fire Brigade                    & mountains, forests, industrial zones        & visual, thermal, multispectral, audio  & high & high\\[6pt]
\hline \\[-6.5pt]
\parbox[t]{12mm}{\multirow{15}{*}{\rotatebox[origin=c]{45}{Helicopter}}} 
& 13 & Land Emergency Situation               & forests, deserts, mountains, rivers         & visual, audio                          & high & high\\
\cmidrule(lr){2-7}
& 14 & Sea Emergency \hspace{4pt}Situation                & islands, coastal regions, seas, lakes       & visual, audio                          & high & high\\
\cmidrule(lr){2-7}
& 15 & Medical Emergency Situation            & urban \& rural \hspace{14pt} remote areas                 & visual, audio                          & high & high\\
\cmidrule(lr){2-7}
& 16 & Assisted/Remote Piloting               & urban \& rural areas                        & visual, audio                          & high & high\\
\cmidrule(lr){2-7}
& 17 & Video Broadcasting \hspace{15pt}of Earth                     & urban \& rural areas \hspace{11pt} of interest            & visual, audio                          & high & low\\
\cmidrule(lr){2-7}
& 18 & Tourist/VIP Aerial Transportation      & tourist areas, remote private areas         & visual, audio                          & high & medium\\
\cmidrule(lr){2-7}
& 19 & Aerial Fire Brigade                    & mountains, forests, industrial zones        & visual, thermal, multispectral, audio  & high & high\\[6pt]
\hline
\end{tabular}
 \begin{tablenotes}[flushleft]
   \item[1]{\fontsize{7.7}{8.8}\selectfont The data rate requirement is specified with respect to the real-time transmission need and data volume.}
   \item[2]{\fontsize{7.7}{8.8}\selectfont The service availability requirement is specified with respect to the criticality of the scenario mission in terms of public safety and human lives.}
 \end{tablenotes}
\end{threeparttable}
\end{table*}

The current work, performed in the context of the ACROSS-AIR activity of the European Space Agency (ESA), 
evaluates the use of broadband SatCom in Rotary Wing Aircraft (RWA), 
i.e., UAVs, UAMs, and helicopters. 
First, 
a set of representative, real-world scenarios
involving RWA with SatCom is presented, 
targeting,
among others, 
public safety,
fast response to emergency situations, 
efficient human transportation,
and 
improved healthcare services. 
These scenarios are extremely challenging and push the technical limits. 
To satisfy their requirements and key performance indicators, 
system modeling is performed for the 
the space, aircraft and communications segments.
For the space segment, active 
LEO, Medium Earth Orbit (MEO) and Geostationary Orbit (GEO)
satellite constellations are employed.
Regarding the aircraft system, well-established RWA of the market are used.
Finally, 
the ever-evolving 
5G Non-Terrestrial Network (5G-NTN) \cite{5g_3gpp} is modeled,
while custom antennas are designed with respect to the specific features of RWA, frequency bands, and constellations.   
The evaluation is based on system-level simulations  
and includes orbit analysis, link budget analysis, blade interference analysis, and End-to-End (E2E) frame transmission.

The contributions of this work are:
(i) this is the first study deploying 5G-NTN SatCom on RWA and examining the blades' interference, to the best of author's knowledge,
and 
(ii) the E2E simulations performed indicate technical challenges and gaps for the full adoption of 5G-NTN on RWA.


\section{Scenarios for Rotary Wing Aircraft with Satellite Communications}
The deployment of SatCom on RWA shall provide new functionalities and improve those that are limited by the use of terrestrial networks. 
Table \ref{tb_scenarios} presents the key parameters of the scenarios considered in the current work. 
Several of the scenarios involve the prevention and handling of emergency situations through surveillance/observation and fast transportation.    
In such occasions, the terrestrial networks may be unstable/down or even unavailable in non-urban regions (e.g., seas, desserts).  
Nevertheless, for increased reliability and performance, the scenarios can support both terrestrial and satellite communications.  
The table also reports the types of data that are generated and transmitted in each scenario.
The majority of data regards multimedia (mainly images and video), 
which impose strict constraints for high data rates, 
especially when the transmission is performed in real time (e.g., in monitoring or broadcasting scenarios). 
Finally, the service availability requirement is qualitatively assessed,
considering the criticality of the scenario for humans. 

\section{System Definition \&  Analysis}
A brief summary on the system definition follows. In particular, the description and system requirements of the satellite constellations, rotary wing aircraft, blades' interference model, and 5G-NTN protocol 
are presented.

\subsection{Satellite Constellations}
The Keplerian elements and payload parameters of the satellites are provided by OHB System AG
and correspond to 
active, operational constellations.  
For the scope of the project,
one GEO, one MEO and two LEO constellations are employed.  
Their Keplerian elements 
are presented in Table \ref{tb_sat1}, 
while the main RF payload parameters for the Ka-band are reported in Table \ref{tb_sat2}.
Additionally, an S-band payload is integrated in the LEO constellations.
This payload includes a patch antenna with a maximum gain of 5dBi that is fed by a transmitter with an output power of 33dBm. 
The S-band antenna possesses a more omnidirectional pattern, while its equivalent noise temperature is equal to 400K. 

\begin{table}[!b]
\vspace{-11pt}
\renewcommand{\arraystretch}{1.1}
\setlength{\tabcolsep}{5.5pt}
\caption{Keplerian Elements of Satellite Constellations}
\vspace{-3pt}
\label{tb_sat1}
\centering
\begin{threeparttable}
\begin{tabular}{l|cccc}
\hline 
                          & \textbf{GEO}  & \textbf{MEO} & \textbf{LEO-1} & \textbf{LEO-2} \\
\hline \hline 
Altitude (km)             & 35786 & 8063 & 1050 & 720  \\ 
Planes (\#)               & 1 & 4 & 12 & 12 \\ 
Inclination ($^{\circ}$)  & 6 &	90, 90, 70, 70  & 89 & 53.5 \\  
RAAN ($^{\circ}$)         & -- & 0, 90, 45, 135 & ($N$-1)$\cdot$15$^*$ &  30 \\ 
Plane Satellites (\#)     & 1 & 6 & 24 & 22 \\ 
Total Satellites (\#)     & 1 & 24 & 288 & 264 \\ 
Configuration             & -- & Star & Star & Delta \\ 
\hline
\end{tabular}
\begin{tablenotes}[flushleft]
   \item[*]{\fontsize{6.3}{7.7}\selectfont For the $N$-th plane.}
  \end{tablenotes}
 \end{threeparttable}
 \vspace{-10pt}
\end{table}
\begin{table}[!b]
\renewcommand{\arraystretch}{1.1}
\setlength{\tabcolsep}{8.0pt}
\caption{RF Ka-Band Payload of Satellite Constellations}
\vspace{-3pt}
\label{tb_sat2}
\centering
\begin{tabular}{l|cccc}
\hline 
                        & \textbf{GEO}  & \textbf{MEO} & \textbf{LEO-1} & \textbf{LEO-2} \\
\hline \hline 
\multirow{2}{*}{Antenna Type}            & Parabolic & \multicolumn{3}{c}{\multirow{2}{*}{Direct Radiating Array}}  \\[-1.7pt] 
                        & Reflector &  &  &   \\ 
Beams (\#)              & 2 &	256  & 64 & 6 \\  
Beam EIRP (dBW)         & 58.1 & 62 & 50 & 40 \\ 
HPBW ($^{\circ}$)       & 0.2 & 2.5 & 4.6 & 2.4 \\ 
G/T (dB/K)              & 12.3 & 10.8 & 5 & 4 \\  
\hline
\end{tabular}
\end{table}

\subsection{Rotary Wing Aircraft}
\begin{figure}[!b]
\vspace*{-23pt}
\centering
\subfloat[DJI's Matrice 30\label{fig_uav1}]{\includegraphics[width=0.465\columnwidth]{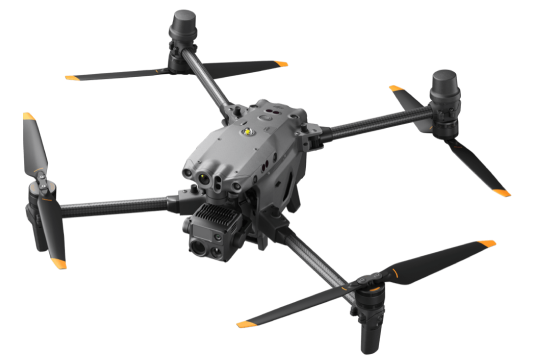}} \hspace{2pt} 
\subfloat[AUS' Alpha 900\label{fig_uav2}]{\includegraphics[width=0.465\columnwidth]{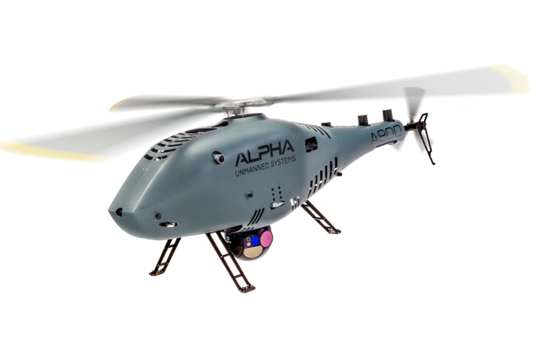}}\\[-10pt]
\subfloat[EHang's EHang 216\label{fig_uam}]{\includegraphics[width=0.42\columnwidth]{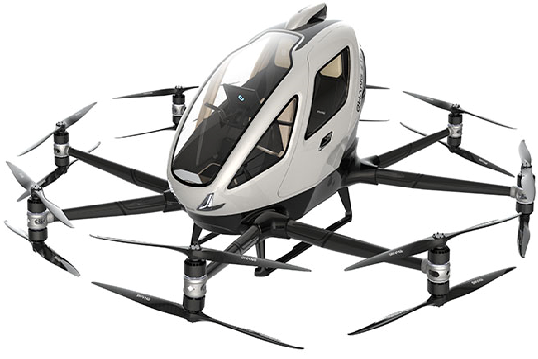}} \hspace{2pt} 
\subfloat[Airbus' H135\label{fig_hel}]{\includegraphics[width=0.465\columnwidth]{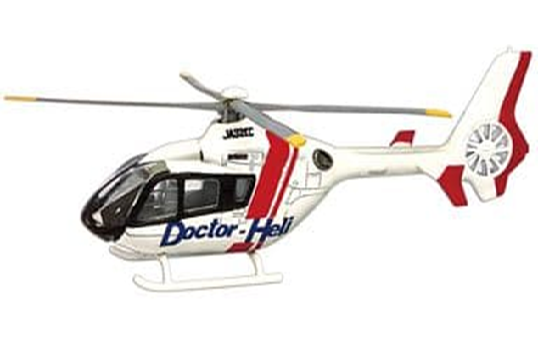}}
\caption{Commercial rotary wing aircraft targeted by the current work.}%
\label{fig_rwas}
\vspace{-3pt}
\end{figure}

The first UAV tested is DJI Matrice 30, illustrated in Fig. \ref{fig_uav1}, which has a high-performance dual camera system with maximum flight time of more than 40 minutes. 
This flight autonomy is ideal for tasks such as public safety and maintenance. Due to its design, the antenna can be located at the center of the main body. Thus, there is no blade interference, 
allowing for a generic performance of SatCom at UAVs.

The second UAV simulated is Alpha 900 from AUS (Fig. \ref{fig_uav2}). 
It is robust and can work over several hours up to 5 Beaufort winds. It is also ideal for inspection tasks, especially in harsh environments, and due to its design, it is expected that the installed antenna will be located below the blades. 
Thus, blades' interference is expected to play an important role.

In the case of UAM, as most of them are still in development phase, the already certified EHang 216 (Fig. \ref{fig_uam}) has been selected. 
It is fully electric with eight propellers, and can be used for transportation between cities, or even for medical transportation. 
An antenna can be installed at the main body, thus avoiding blades' interference
and allowing again
for a generic assessment.

For the helicopter scenarios, the proven Airbus H135 was selected (Fig. \ref{fig_hel}), which can be used for a multitude of tasks, from passenger transportation to emergency medical services and law enforcement. It is compact and maneuverable, with four blades at its main rotor. The necessary SatCom antenna is expected to be installed at a point where blades' interference exists. 
Also, due to its dimensions, mechanically steerable antennas for communication with GEO satellites is possible.

Table \ref{tb_aircraft} summarizes the specifications of RWA's antennas.
These antennas have been modelled to satisfy the system requirements and features of RWA.

\begin{table}[!b]
\renewcommand{\arraystretch}{1.1}
\setlength{\tabcolsep}{1.7pt}
\caption{Antenna Characteristics of Rotary Wing Aircraft}
\vspace{-3pt}
\label{tb_aircraft}
\centering
\begin{tabular}{l|cccc}
\hline 
                          & \textbf{UAV-1}  & \textbf{UAV-2} & \textbf{UAM} & \textbf{HELI} \\
\hline \hline 
Manufacturer      & DJI & AUS & EHang & Airbus  \\ 
Model             & Matrice 30 & Alpha 900 & EHang 216 & H135  \\
\hline 
\multirow{2}{*}{Antenna Type}          &  Patch  & Patch  & Phased  & Parabolic /  \\[-1.7pt]
         &   Array &  Antenna &  Array &  Phased Array \\ 
Freq. Band       &  Ka-Band & S-Band & Ka-Band & Ku- / Ka-Band  \\
Bandwidth (MHz)        & 30 & 30 & 400 & 36 / 400 \\
Beamwidth ($^{\circ}$)        & 26.2 & 89.8 & 3.2--4.4 & 2 / 3.2--4.4 \\
Max Gain (dBi)     & 17.33 & 5.15 & 36.26 & 41.4 / 36.26 \\  
Position on RWA    &  main body &  under blades & main body & under blades \\  
\hline
\end{tabular}
\end{table}

\subsection{Blade Interference in Communication Link}
\begin{figure}[!t]
    \centering
    \includegraphics[trim= {0 0 0 2}, clip, width=0.87\columnwidth]{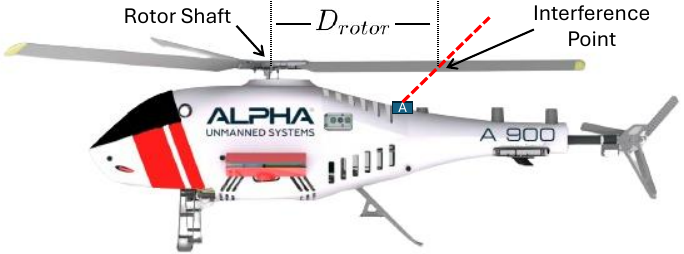}
    \vspace{-5pt}
    \caption{Modeling of the blades' interference in the SatCom link of RWA.}
    \label{fig_blades_meth}
   \vspace{-15pt}
\end{figure}

To calculate the interference of the RWA's blades in the link,
a simplified model has been developed. 
Without loss of generality, it is assumed that the antenna, placed under the blades,
points as shown in Fig. \ref{fig_blades_meth}.
The point where the antenna direction meets the blade is defined as ``interference point'' and depends on the antenna's position and the elevation angle.

Let $t_{start}$ and $t_{stop}$ be the timestamps that the blade 
starts and stops, respectively, 
to pass from the interference point. 
Hence, the interference time of each rotor's blade is $t_{int} = t_{stop} - t_{start}$.
In the time interval $t_{int}$, 
the blade is rotated approximately 
by distance equal to its width $W$,
creating an angle $\phi$.
$W$ is the arc of $\phi$,
while the radius of the circle formed,
labeled as $D_{rotor}$, 
is equal to 
the distance between the interference point and the rotor shaft (circle center). 
The angle $\phi$ is calculated as follows:
\begin{equation}
    W = 2 \pi \cdot D_{rotor} \cdot (\phi / 360) \Rightarrow \phi = 360 W / (2 \pi \cdot D_{rotor})
\end{equation}

Let $R_{dms} = 0.006 \cdot R_{rpm}$
be the blades' rotational speed in deg/ms. 
Assuming $t_0 = t_{start}$,
the interference time of each blade, i.e., 
$t_{int}$, is calculated as follows:\\[-8pt]
\begin{equation}
    t_{int} = \phi/R_{dms}
\end{equation}

Let $T_{rot} = 360/R_{dms}$
be the blades' rotation time,
and $N_{bld}$ be the number of blades. 
The total non-interference time in one rotation is calculated as follows:\\[-8pt]
\begin{equation}
    T_{lnk} = T_{rot} - N_{bld} \cdot t_{int}
\end{equation}

Hence, the time interval between two interferences, i.e., the continuous link time, is calculated as follows: \\[-8pt]
\begin{equation}
    t_{lnk} = T_{lnk}/N_{bld}
\end{equation}

\begin{figure}[!t]
\centering
\hspace{-9pt}\subfloat[Alpha 900\label{fig_bld1}]{\includegraphics[ width=0.52\columnwidth]{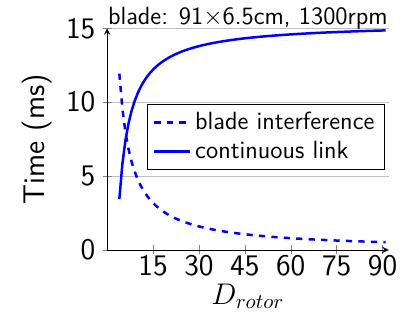}}%
\hspace*{-4pt}
\subfloat[H135\label{fig_bld2}]{\includegraphics[ width=0.52\columnwidth]{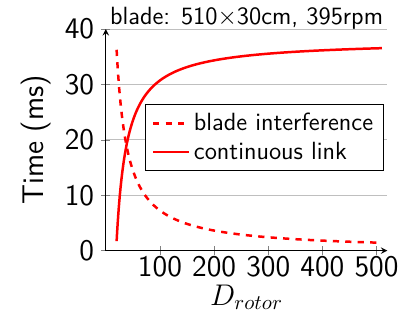}}
\caption{Blades' effect in the SatCom link of RWA.}%
\label{fig_blades_results}
\end{figure}

Fig. \ref{fig_blades_results} illustrates the results of the blade analysis  
for Alpha 900 and H135.  
Alpha 900 
has a 3.2$\times$ higher rotational speed than H135. 
Thus, it has smaller $t_{int}$ and $t_{lnk}$. 
It is also observed that $t_{int}$ is reduced as $D_{rotor}$ increases. 
Finally, it is worth mentioning that 
due to aerodynamics and the relative attitude of the RWA when moving, 
in reality, 
the projected blades' shadowing is smaller, and $D_{rotor}$ is larger than modeled, i.e., a bit smaller  interference times are expected. 

\subsection{5G-NTN Satellite Communications}

5G-NTN is being developed by 3GPP based on 5G New Radio (NR). 
In the current work,
the Physical Downlink Shared Channel (PDSCH),  
Physical Uplink Shared Channel (PUSCH), 
and NTN channels 
are modeled in the system-level simulations.  
Table \ref{tb_5gntn} reports the operating bands and the channel arrangement of 5G-NTN, 
while Table \ref{tb_5gframe} summarizes key parameters about the 10ms 5G frame and the bandwidth.

\section{System Modeling \& Simulation}

For simulating the orbits and the satellite constellations,
the \emph{Ansys STK} tool is employed. 
For the 5G-NTN modeling, link budget calculations and E2E simulations, 
the \emph{MATLAB} tool is used. 
Regarding the scenarios, five of them from Table \ref{tb_scenarios} are selected,
which impose increased technical challenges and strict requirements 
for data rate and service availability. 
Moreover, realistic flight duration and routes are selected 
based on the aircraft autonomy and the mission's objectives.


\begin{table}[!t]
 \vspace{-6pt}
\renewcommand{\arraystretch}{1.1}
\setlength{\tabcolsep}{2.6pt}
\caption{Operating Bands and Channel Arrangement 
 of 5G-NTN}
\vspace{-3pt}
\label{tb_5gntn}
\centering
\begin{threeparttable}
\begin{tabular}{c c| >{\centering\arraybackslash}m{1.5cm} >{\centering\arraybackslash}m{1.5cm} >{\centering\arraybackslash}m{1.5cm} >{\centering\arraybackslash}m{1.5cm}}
\hline 
\multicolumn{2}{c|}{\textbf{Freq. Band}}  & \textbf{UL Freq. (GHz)} & \textbf{DL Freq. (GHz)} & \textbf{Channel \hspace{11pt} BW (MHz)} & \textbf{Subcarrier Spac. (kHz)}\\
\hline 
\hline 
\multirow{3}{*}{NTN-FR1}  & n254 & 1.61--1.63 & 2.48--2.50 & 5/10/15         & \multirow{3}{*}{15/30/60} \\
                          & n255 & 1.63--1.66 & 1.53--1.56 & 5/10/15/20     & \\
                          & n256 & 1.98--2.01 & 2.17--2.20 & 5/10/15/20/30 & \\
                          \hline 
\multirow{3}{*}{NTN-FR2}  & n510 & 27.50--28.35 & 17.30--20.20  & \multirow{3}{*}{50/100/200/400}         & \multirow{3}{*}{60/120} \\
                          & n511 & 28.35--30.00 & 17.30--20.20 &      & \\
                          & n512 & 27.50--30.00 & 17.30--20.20 &  & \\                          
\hline
\end{tabular}
\begin{tablenotes}[flushleft]
   \item[*]{\fontsize{6.3}{7.7}\selectfont \underline{Source}: 3GPP TS 38.101-5 Version 18.5.0 Release 18.}
  \end{tablenotes}
 \end{threeparttable}
 \vspace{-7pt}
\end{table}

\begin{table}[!t]
\renewcommand{\arraystretch}{1.1}
\setlength{\tabcolsep}{4.5pt}
\caption{5G-NTN Frame Structure and Bandwidth Configuration}
\vspace{-3pt}
\label{tb_5gframe}
\centering
\begin{threeparttable}
\begin{tabular}{>{\centering\arraybackslash}m{1.45cm}| >{\centering\arraybackslash}m{1.45cm} >{\centering\arraybackslash}m{1.45cm} >{\centering\arraybackslash}m{1.45cm} >{\centering\arraybackslash}m{1.45cm}}
\hline 
\textbf{Subcarrier Spac. (kHz)} & \textbf{Slots per Frame (\#)} & \textbf{Slot Length (ms)} & \textbf{Resource Blocks (\#)} & \textbf{Channel BW (MHz)} \\
\hline 
\hline 
15  & 10  & 1     & [25, 160] & [5, 30]\\ 
30  & 20  & 0.5   & [11, 78] & [5, 30]\\ 
60  & 40  & 0.25  & [11, 264] & [10, 200]\\ 
120 & 80  & 0.125 & [32, 264] & [50, 400]\\ 
\hline
\end{tabular}
\begin{tablenotes}[flushleft]
   \item[*]{\fontsize{6.3}{7.7}\selectfont \underline{Source}: 3GPP TS 38.211 Version 18.2.0 Release 18.}
  \end{tablenotes}
 \end{threeparttable}
 \vspace{-4pt}
\end{table}

\begin{table*}[t]
\vspace{-8pt}
\renewcommand{\arraystretch}{1.15}
\setlength{\tabcolsep}{4pt}
\caption{Orbit \& Link Budget Analysis for Real-World Scenarios with Rotary Wing Aircraft and Satellite Communications}
\vspace{-3pt}
\label{tb_satres}
\centering
\begin{threeparttable}
\begin{tabular}{cccccccccccccc}
\hline                               
\multirow{2.3}{*}{\textbf{Scen.}}
&
& 
\multirow{2.3}{*}{\textbf{Satellite}} & 
\multirow{2.3}{*}{\textbf{Flight}}    &
                                      &
\multicolumn{2}{c}{\textbf{Elevation Angle ($\mathbf{^{\circ}}$)}}  & 
\multicolumn{2}{c}{\textbf{Doppler Shift (kHz)}} & 
\multicolumn{1}{c}{\textbf{Loss (dB)\tnote{3}}} & 
\multicolumn{2}{c}{\textbf{CNR (dB)}} & 
\multicolumn{2}{c}{\textbf{CNR$'$ (dB)\tnote{4}}} \\
\cmidrule(lr){6-7}\cmidrule(lr){8-9}\cmidrule(lr){10-10}\cmidrule(lr){11-12}\cmidrule(lr){13-14}
\textbf{ID} & 
\textbf{RWA} &  
\textbf{Constel.} & 
\textbf{Time\tnote{1}} & 
\textbf{Link / Band} &
\emph{Range} & \emph{Avg.} &
\emph{Range} & \emph{Avg.} &
 \emph{Average} &
\emph{Range} & \emph{Avg.} &
\emph{Range} & \emph{Avg.} \\
\hline \hline 
6 & UAV-1 & LEO-2 & 0.5h  & UL / Ka-Band  & [35.6, 89.3] & 56.7 & [8, 525]  & 346 & 181 & [-64.1, 0.5] & -17.6 & [-61.2, 3.5] & -14.6 \\ 
7 & UAV-2 & LEO-1 & 2h    & UL / S-Band  & [40.3, 89.3] & 58.3   & [0.6, 35] & 24 & 161  & [-3.5, 4.7]   & 1.1 & [4.2, 12.5] & 8.8 \\   
11 & UAM  & MEO   & 1.45h & DL / Ka-Band  & [39.2, 85.6] & 61.6  & [11, 113] & 68  & 198 & [2.6, 7.9]    & 6.1  & [8.5, 14] & 12.1 \\  
15a & HELI  & LEO-1 & 2h    & UL / Ka-Band  & [39.5, 89.1] & 59.1  & [10, 466] & 305 & 189 & [-24.7, 7.6]  & 0.1 & [-21.6, 10.7] & 3.2 \\  
15b & HELI  & GEO & 2h      & UL / Ku-Band\tnote{2}  & [18.9, 23.8] & 21.4  & [20, 21]  & 20.5 & 210 & [5.1, 15.8]   & 13.6 & -- & -- \\  
19 & HELI  & LEO-2 & 2.5h  & DL / Ka-Band  & [34.6, 88.7] & 54.5  & [10, 351] & 242 & 180 & [6.4, 28.9]   & 22.9 &  -- & -- \\  
\hline
\end{tabular}
\begin{tablenotes}[flushleft]
    \item[1]{\fontsize{6.3}{7.7}\selectfont The satellite access percentage is 99.9\% in all scenarios.}
    \item[2]{\fontsize{6.3}{7.7}\selectfont Ku-Band is not applicable to 5G-NTN, however, it has been selected for comparison purposes.}
    \item[3]{\fontsize{6.3}{7.7}\selectfont The propagation loss includes free space loss and Earth-space losses based on ITU P.618 model (heavy rain also appears during the flights).}
    \item[4]{\fontsize{6.3}{7.7}\selectfont CNR$'$ is calculated for reduced channel bandwidth compared to CNR (lowest bandwidth allowed in the frequency band for scenarios 6, 7, 15a).}
   \end{tablenotes}
 \end{threeparttable}
\vspace{-7pt}
\end{table*}

\begin{figure*}[!t]
\centering
\subfloat[Scen. ID = 6\label{fig_5g1}]{\includegraphics[trim= {91 267 91 269}, clip,width=0.33\textwidth]{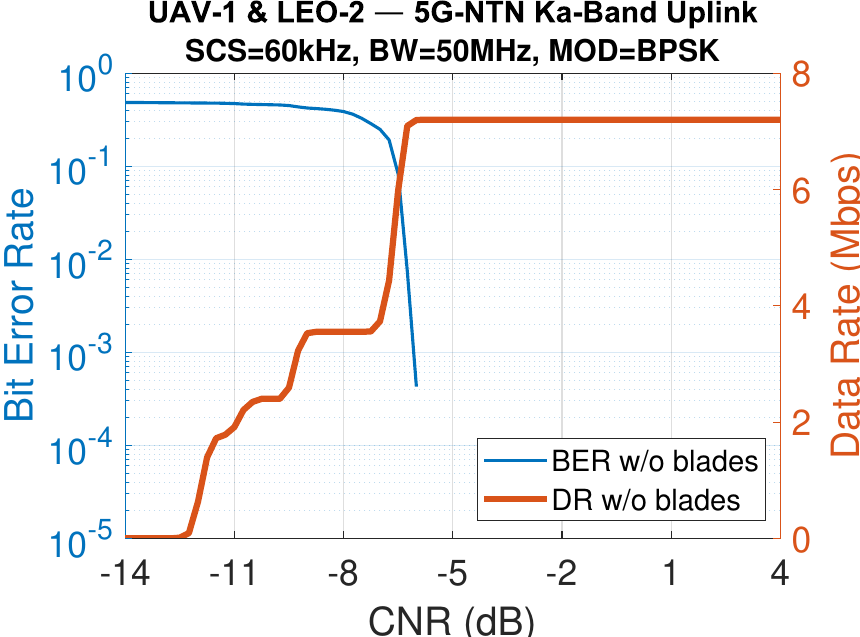}}\hfill
\subfloat[Scen. ID = 7\label{fig_5g2}]{\includegraphics[trim= {91 267 91 269}, clip,width=0.33\textwidth]{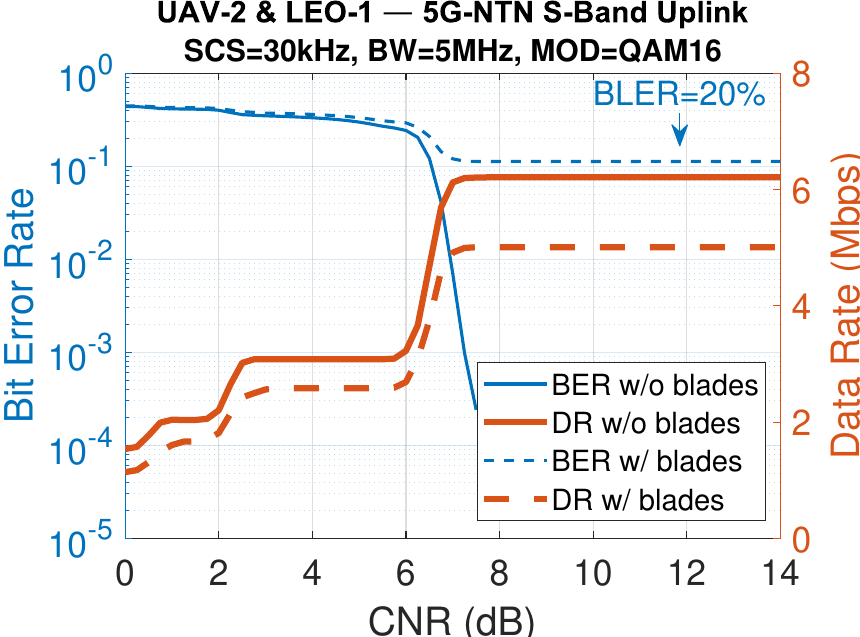}}\hfill
\subfloat[Scen. ID = 11\label{fig_5g3}]{\includegraphics[trim= {91 267 91 269}, clip,width=0.33\textwidth]{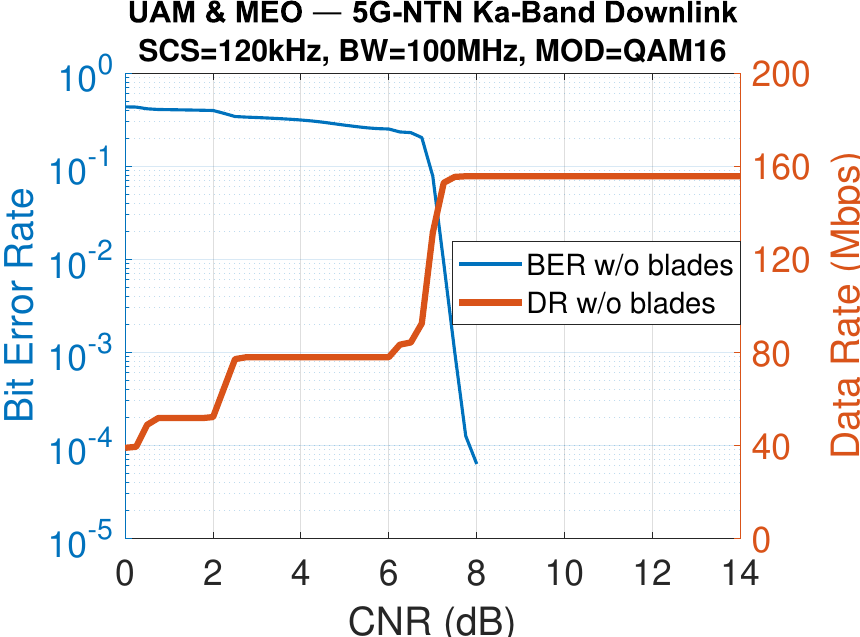}}\\[-3pt]
\subfloat[Scen. ID = 15a\label{fig_5g4}]{\includegraphics[trim= {91 267 91 269}, clip,width=0.33\textwidth]{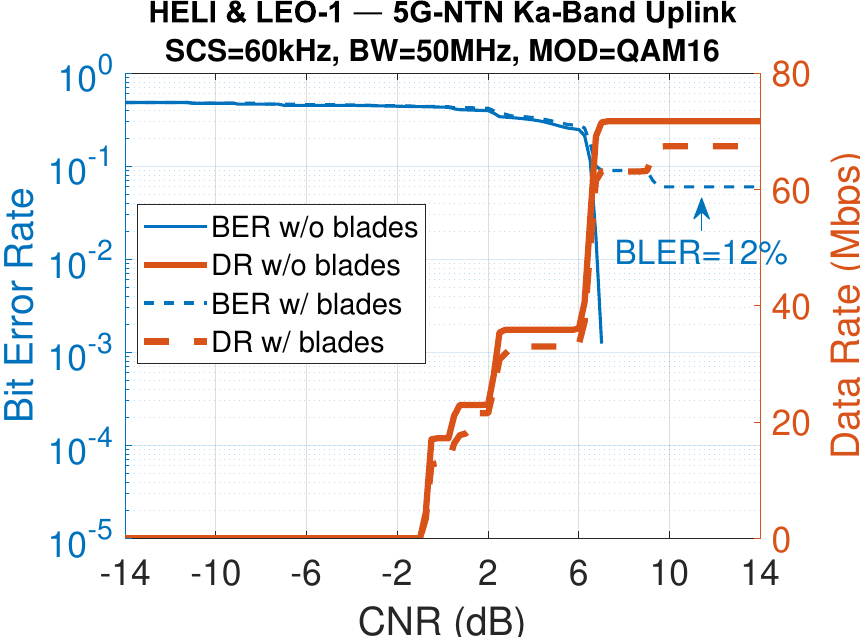}}\hfill
\subfloat[Scen. ID = 15b\label{fig_5g5}]{\includegraphics[trim= {91 267 91 269}, clip,width=0.33\textwidth]{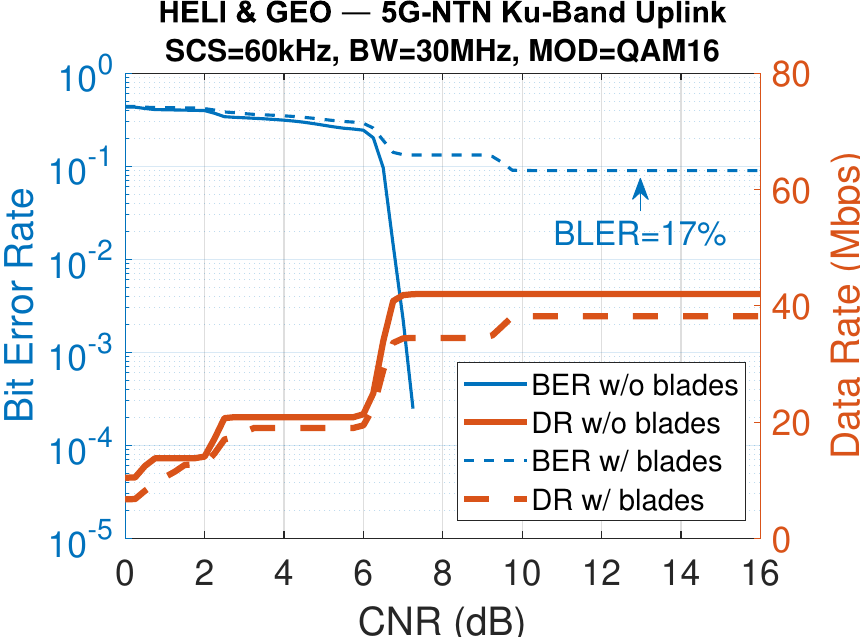}}\hfill
\subfloat[Scen. ID = 19\label{fig_5g6}]{\includegraphics[trim= {91 267 91 269}, clip,width=0.33\textwidth]{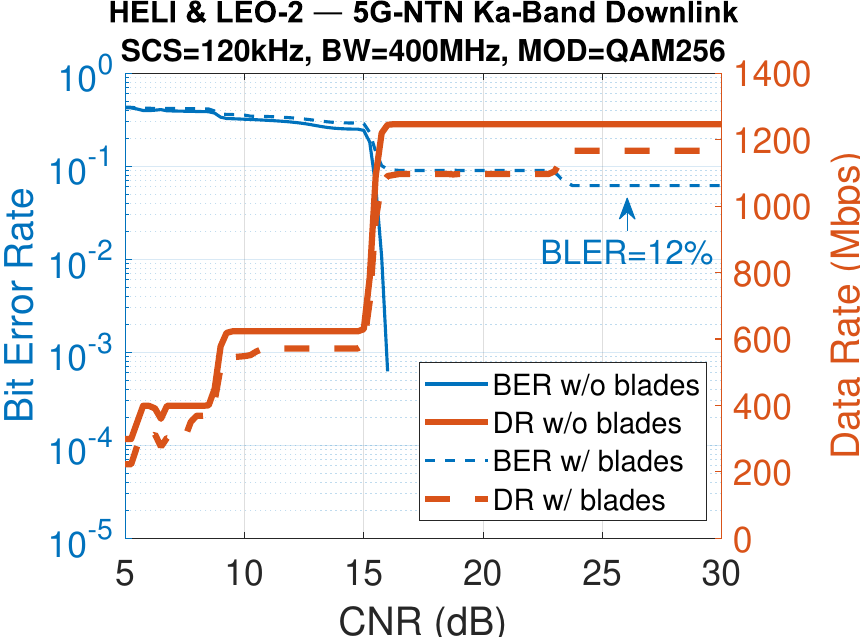}}
\caption{End-to-end 5G-NTN simulations for the RWA scenarios, tailored to the CNR range of each flight (Table \ref{tb_satres}).}%
\label{fig_5g}
\vspace{-3pt}
\end{figure*}

Table \ref{tb_satres} reports the results of the orbit and link budget analysis.
A satellite handover is performed when the elevation angle drops below 
40$\mathbf{^{\circ}}$--35$\mathbf{^{\circ}}$.
The propagation loss
includes free space loss and troposphere impairments (e.g., rain and cloud attenuation).
Both the Doppler shift and the propagation loss are within the expected range
considering the orbit and the frequency band. 
In case the Carrier-to-Noise Ratio (CNR) is low,
the analysis is also performed for lower bandwidth,
where CNR$'$ is calculated. 
The other solution would be to increase the transmit power,
which is not always feasible,
as it may require re-design like in the case of small UAVs.

Fig. \ref{fig_5g} presents the Bit Error Rate (BER) and data rate curves 
of the 5G-NTN simulations corresponding to the scenarios. 
For the 5G waveform configuration, an extensive exploration of the subcarrier spacing, channel bandwidth, modulation order, and transport block size has been carried out. 
The selected configurations 
aim to provide low error rates within the CNR range of the scenarios,
while also taking into account the available bandwidths of each frequency band (see Table \ref{tb_5gntn}).
The simulations have been performed for 100 5G frames. 

First, it is observed that the blades have a significant impact on BER,
with the average $t_{int}$ being 1.6ms and 2--3.2ms
for UAV-2 and HELI, respectively. 
This is justified by the fact that $t_{int}$ \raisebox{0.8pt}{{\tiny\textgreater}} $t_{slot}$ (see Table \ref{tb_5gframe}), namely entire slots (transport blocks) are lost. 
For example, in case of Scenario 7, where $t_{int}$$=$1.6ms and $t_{lnk}$$=$13.9ms on average, 
approximately 3 slots are lost and then 28 are passed 
in a recurring pattern.
Namely, BER lies around 10$^{-\text{1}}$ because \raisebox{0.8pt}{{$\sim$}}10\% of the slots are entirely lost.
Some data from the first and last passed slots is also lost.  
Nevertheless, all these errors are burst (periodical) in nature and not random,
and they mainly regard the lost slots. 
Thus, the Block Error Rate (BLER) is 
\raisebox{0.8pt}{{$\sim$}}10--20\%, i.e., for every 10 blocks, around 8--9 of them are transmitted without errors. 


Second, taking into account the average data rates of the UAM and HELI flights,
high-resolution video can be transmitted, especially when using the H.265 compression standard.
Even in the case of UAVs, low-resolution video can be streamed 
during the surveillance and observation missions. 
Regarding the blade interference, 
it results in fewer blocks being decoded at the receiver, and thus, the data rate is decreased. 
Third, the low performance of Scenario 6 is mainly attributed to the UAV-1's patch array,
which has a fixed radiation pattern, contrary to HELI's phased array that has a steerable radiation pattern.
As a result, 
this antenna cannot adjust its steering to the passing LEO satellites.
When comparing Scenario 15a (LEO, Ka-band) and Scenario 15b (GEO, Ku-band), 
the latter provides better CNR due to its antenna (more dbW and better max. gain). 
Finally, it is worth mentioning that the heavy rain has a significant negative impact on the performance. 

\section{Discussion}
The ACROSS-AIR project has investigated system-wise the use of SatCom for RWA.  
Through this way, 
key results and outcomes have been produced, 
which led to the identification of several 
technical and non-technical challenges \& open issues. 
These points cover all the necessary components for the successful implementation of the SatCom operations,
covering 
the space, aircraft, and telecommunications segments.
Subsequently, some significant challenges are briefly presented:

-- \emph{\underline{Small vs Large RWA}}:
        RWA operate at relatively low heights compared to fixed wing aircraft. Thus, the distance from the aircraft to the satellite is larger and the UL signal is not capable to transmit real-time video; usually, telemetry data in real time is the absolute maximum. 
        For a large RWA, installing several patch antennas and large amplifiers (in the case of LEO connections) will possibly solve the problem. 
        Additionally, to communicate with multi-orbit constellations, i.e., MEO and LEO, and due to the existence of handovers, the development of conformal antennas not affecting the aircraft's airworthiness is required.
        For small RWA, this is almost impossible with the current technology. 
        The use of novel materials with improved size, weight, and power is required to comply with the strict  space and power constraints of RWA.
        Namely, new components, such as more powerful amplifiers with smaller dimensions or denser patch antennas, are needed.

        
        
   --  \emph{\underline{Paradigm Shift in E2E Architecture}}:
        Future satellite constellations are in the pipeline from various private entities, while in Europe, the IRIS$^2$ constellation is being designed. In several of these cases, there are proprietary protocols, or the terminals can connect with specific types of satellites. The use of 5G-NTN, as well as the future versions, i.e., B5G and 6G, solve the problem to some extent. 
        However, and perhaps for some particular operations, such as S\&R or for medical emergencies, a different multi-layered architecture has to be developed and embedded in the next SatCom systems. This architecture could be able to operate with any satellite system available,
        like roaming in mobile telephony. 
        
   --  \emph{\underline{Design of Future Satellites}}:
        Given the fact that it may be difficult for UAVs and UAMs 
        to increase their UL capabilities beyond a threshold, the other end of this link may be able to be changed. More powerful but compact satellites, especially for LEO, can be explored. Current emerging technologies, such as AI and SDR, can help on that, but also larger receptors have to be designed. In the case of smallsats, deployable structures that can be folded several times during launch pave the way for such a capability.
        
    --  \emph{\underline{Communication Protocols Enrichment}}:
        Although the 5G-NTN protocol is many steps ahead from the early releases of the 3GPP guidelines, there are still many open points for the use of SatCom in the cases of RWA. For example, a consistent way to counteract the intermissions created by the rotation of the blades, or specific adaptations of the protocol to allow for more efficient use of UL for the smaller RWA, are necessary.

All these challenges are to an extent generic,   and they can be considered as the overall goals to be achieved for incorporating the efficient use of SatCom for RWA. Apparently, each of these goals has to be further analysed into separate but interconnected technological building blocks, while related roadmaps for the next 5-10 years have to be developed.

\vspace{7pt}

\section{Conclusion}
In this work, 
an E2E analysis for the performance of 5G-NTN SatCom for manned and unmanned RWA was presented. 
Various realistic scenarios were examined, and  system requirements were reported. 
First, an orbit \& link budget analysis took place 
to extract key system parameters such as the elevation angles, Doppler shifts, propagation losses and CNRs. 
Subsequently, BER, BLER and data rate results
were presented from 
5G-NTN simulations corresponding to a diverse set of scenarios
with various aircraft, orbits, links, and frequency bands. 
In summary, the small UAVs
challenge the antenna design when targeting high data rates and low error rates. 
The blades also constitute a challenging issue, as they increase BER to 10$^{-\text{1}}$.
However, these errors are burst/periodical in nature and not random, 
and thus, 
BLER lies around 10--20\%, i.e., 
8--9 out of 10 transport blocks are successfully transmitted without errors. 
The theoretical data rates of 5G are achieved 
by the large RWA (UAMs and helicopters) in clear sky conditions,
exploiting the capabilities of their phased array antennas.


\vspace{7pt}

\section{Acknowledgement}
This work has been supported by the European Space
Agency (ESA) via the funded activity ACROSS-AIR:  
``Advanced Broadband SatCom Solutions for Rotary Wing Aircraft'' with reference number 4000141490/23/NL/MM/gm (\href{https://connectivity.esa.int/projects/acrossair}{\textcolor{blue}{https://connectivity.esa.int/projects/acrossair}}).
The views of the authors do not necessarily reflect the views of ESA.

\vspace{7pt}

\bibliographystyle{IEEEtran}
\bibliography{REFERENCES.bib}

\end{document}